\documentclass[twocolumn]{webofc}

\usepackage[varg]{txfonts}   
\usepackage{hyperref}
\usepackage{url}
\hypersetup{colorlinks=true,citecolor=blue,urlcolor=blue,linkcolor=blue}
\begin{document}
\title{Preliminary experimental investigation on the interaction of a subaqueous dune like granular structure with a turbulent open channel flow}
%
%

\author{\firstname{Durbar} \lastname{Roy}\inst{}\thanks{\email{durbarroy681@gmail.com}} \and
        \firstname{Ikbal} \lastname{Ahmed}\inst{} \and
        \firstname{Abdul} \lastname{Hakkim}\inst{} \and
        \firstname{Rama} \lastname{Govindarajan}\inst{}
}

\institute{International Center for Theoretical Sciences, Tata Institute of Fundamental Research, Bengaluru 560089, India}

\abstract{
We study the interaction of a subaqueous dune like granular structure with a
turbulent open channel flow experimentally using
optical diagnostics in the Shields and Froude parameter space ($0.05{\leq}{\theta}{\leq}0.32$, $0.1{\leq}Fr{\leq}0.4$).
Interactions between the turbulent flow and the
granular structure give rise to transient erosion-deposition dynamics leading to various types of particle transport. The subaqueous structures in the channel bed evolves due to shear-stress-induced erosion, gravity-driven deposition, and subsequent particle transport. We study the centroid motion and the granular structure shape evolution. At lower end of our ${\theta}-Fr$ parameter space, we observe no erosion and the structure remains at rest.
We show that the critical Shields number (${\theta}_{cr}$) is of the order of $0.1$ beyond which erosion starts to occur.
At intermediate values of ${\theta}$ and $Fr$ we observe slow erosion, resulting in a rigid body motion of the granular structure
without significant shape deformation.
Higher values of ${\theta}$ and $Fr$ causes vortex formation at the upstream of the dune resulting in stronger erosion, rapid shape deformation and relatively higher translation velocity of the centroid.
}
\maketitle
\section{Introduction}
\label{intro}
\begin{figure}[h]
\centering
\includegraphics[width=7cm,clip]{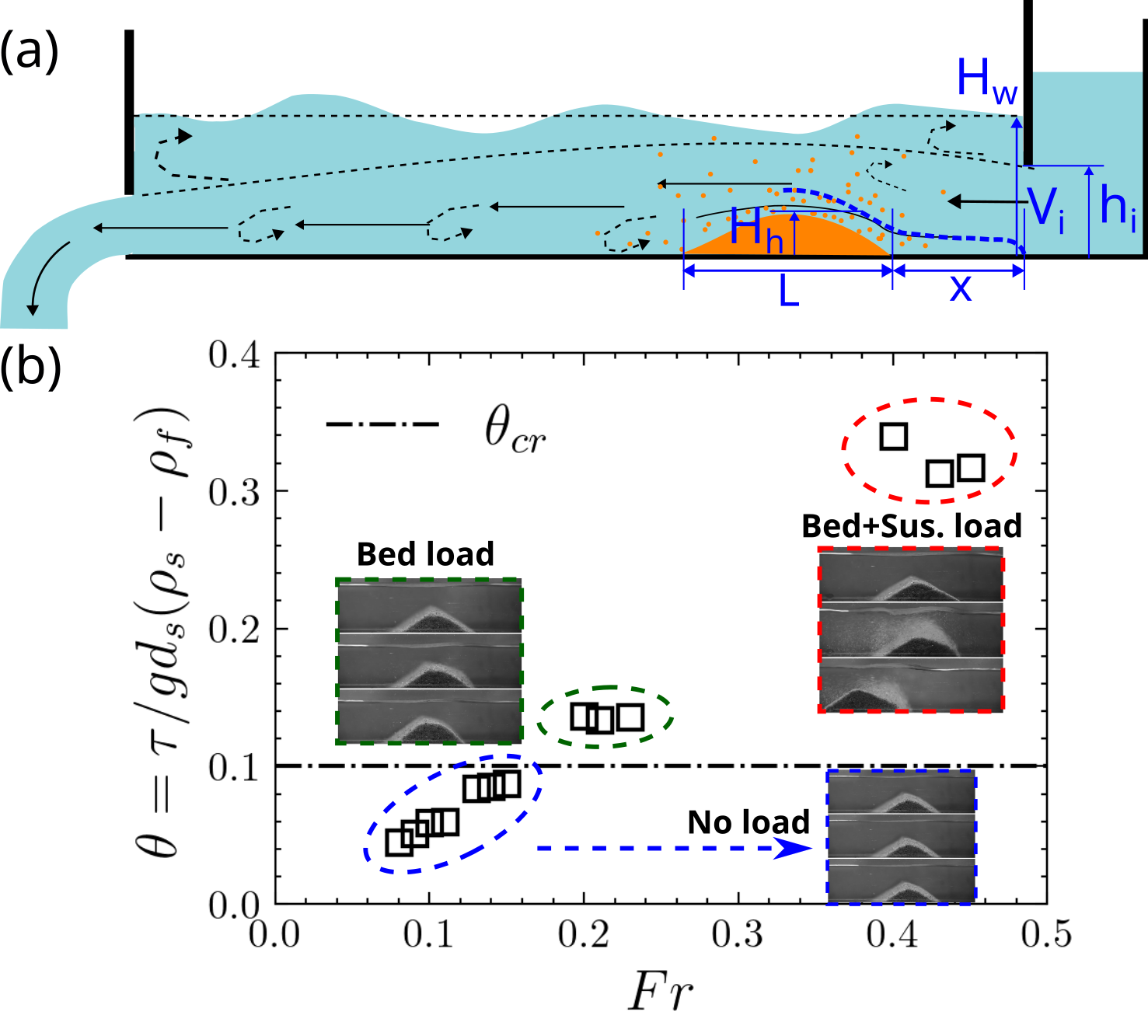}
\caption{(a) Schematic depicting the problem geometry: Turbulent fluid flow in an open channel flow interacting with a subaqueous dune like structure. (b) Experimental regime map in ${\theta}-Fr$ (Shields and Froude number) space showing the various kinds of particle transport and dune movement/deformation.}
\label{fig-1}       
\end{figure}

\begin{figure*}[h]
\centering
\vspace*{1cm}       
\includegraphics[width=15cm,clip]{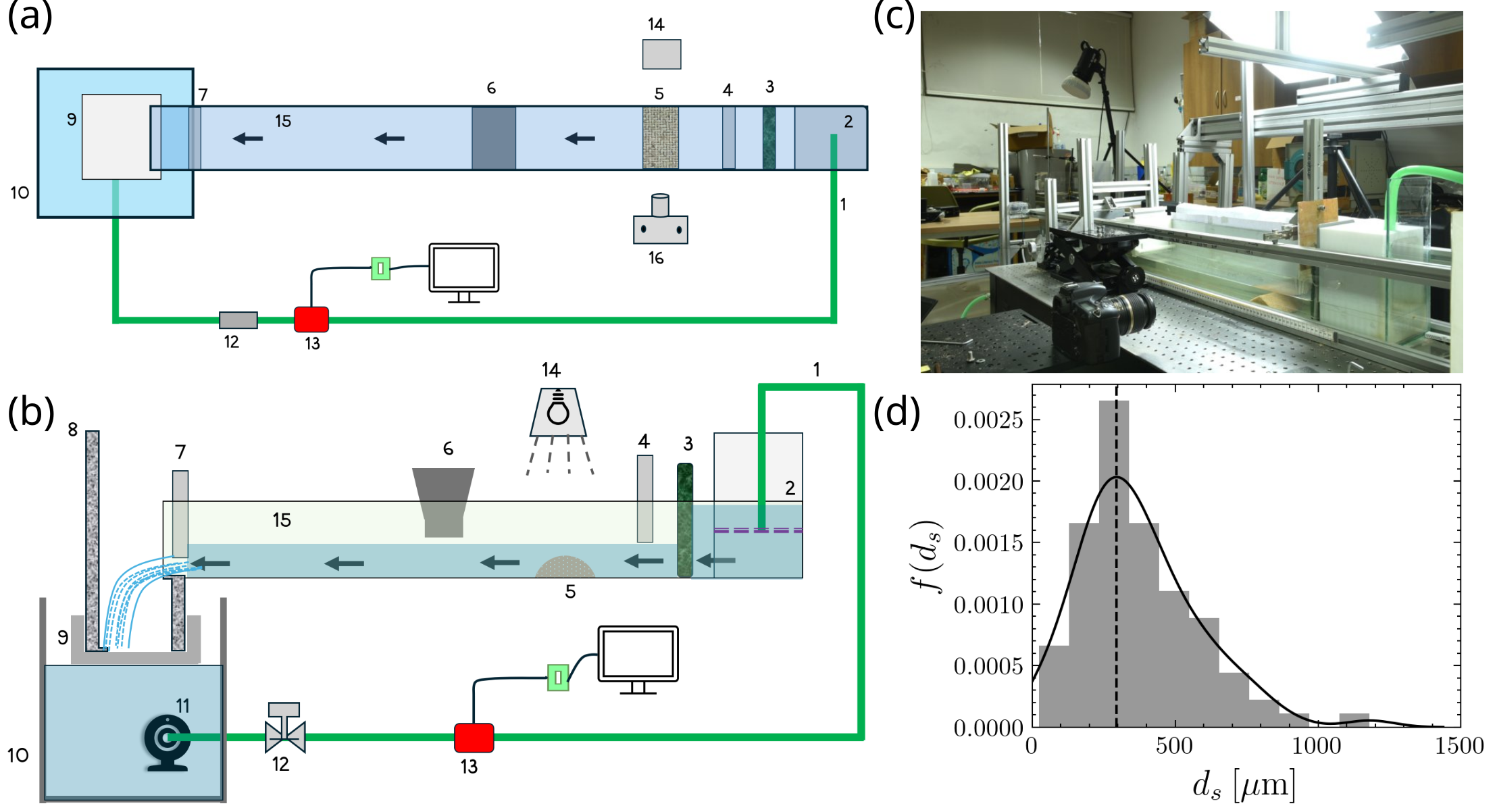}
\caption{Schematic depicting the (a) top, (b) side view of the experimental setup. The components numbered are 1: Pipe, 2: Inlet section, 3: Flow conditioner, 4: Inlet gate, 5: Sand heap, 6: Sand dispenser funnel, 7: Exit gate, 8: Exit guide duct, 9: Strainer, 10: Storage tank, 11: Submersible pump, 12: Flow control valve, 13: Flow meter, 14: Light source, 15: Flume/channel, 16: Camera. (c) Image of actual experimental setup. (d) Probability density function depicting sand size distribution in microns.}
\label{fig-2}       
\end{figure*}
The fascinating structures occurring from geomorphological processes
in fluvial systems \cite{balmforth2008geomorphological} have captivated the imagination of humans for eons for both practical applications (geology, hydraulic engineering) \cite{engelund1970instability} and curiosity driven science. For example, Albert Einstein was known to be interested in meandering rivers and Baer-Babinet law \cite{einstein1926cause,balla2007influence}. Inspired by such structures like meandering rivers, river networks \cite{rodriguez2001fractal}, desert dunes \cite{bagnold2012physics,charru2013sand}, we study the erosion-deposition dynamics of granular particles in the context of subaqueous geophysical processes.
Majority of the previous studies were done in the context of sand dunes \cite{bagnold2012physics}, however similar studies in the context of subaqueous structure are relatively sparse \cite{franklin2011subaqueous,charru2012subaqueous,groh2008barchan,hersen2002relevant}. Interactions between a turbulent fluid flow and a granular bed give rise to various patterns like ripples, dunes and chevrons due to instabilities \cite{andreotti2012bedforms,charru2013sand}.  
Some of the previous studies investigated subaqueous barchan dunes experimentally in high depth configurations
\cite{groh2008barchan,hersen2002relevant,groh2009attractive,du2015dune,bacik2021stability,assis2025evolving}. In the present work, using high-fidelity optical diagnostics, we study the turbulent flow interactions with a granular structure at finite water depth for various inlet conditions typically found in natural systems like streams, rivers and continental shelf to name a few .
Fig. \ref{fig-1}(a) shows a schematic representation of the problem geometry, consisting of a steady (constant volume flow rate: $Q$) turbulent flow with an inlet velocity scale $V_i$, inlet gate height $h_i$, a free surface water height $H_w$ interacting with a subaqueous granular structure with an initial condition of an isosceles triangular cross section profile of a given horizontal (length of the structure: $L$) and vertical length scales (height of the structure: $H_h$) respectively throughout the entire width ($w$) of the channel. The velocity scale $V_i$ is related to the flow rate $Q$ and inlet gate height $h_i$ as $Q{\sim}wh_iV_i$
where $w$ is the constant width of the channel. The leading edge of the dune is located at a distance $x$ from the inlet.
The discharge per unit width $q$ scales as $q={Q}/{w}{\sim}h_iV_i$ and hence the inflow velocity scales as $V_i{\sim}{q}/{h_i}$. The Froude number $Fr$ defined as $Fr={V_i}/{\sqrt{gH_w}}$, where $g$ is the acceleration due to gravity, measures the ratio of inertia force to gravity force is one of the important non-dimensional numbers used to characterize the current flow to study the effects of free surface.
On using the scale of $V_i$ in the definition of $Fr$, we have $Fr={q}/{h_i\sqrt{gH_w}}$. The ratio of inertia to viscous force is characterized by the Reynolds number which is significantly high owing to the turbulent nature of the channel flow. The inlet Reynolds number of the flow is defined as $Re_i={V_ih_i}/{\nu}$, where $\nu$ is the kinematic viscosity of fluid. It is important to note that the product $V_ih_i=q$ which is a constant, hence $Re_i=q/{\nu}$ is a constant for a fixed $q$.
The other non-dimensional number that characterizes the degree of erosion is the Shields number (${\theta}$) which is a non-dimensional form of
shear stress defined as ${\theta}={\tau}/({\rho}_s-{\rho}_f)gd_s$, where ${\tau}$, ${\rho}_s$, ${\rho}_f$, $d_s$ is the turbulent shear stress,
density of sand grains, density of fluid and size of sand grains respectively. Computing the turbulent stress ${\tau}$ acting on the subaqueous granular structure is non-trivial and requires sophisticated theoretical and experimental techniques. In the current preliminary study we estimate the turbulent shear stress from the scale of shear stress at the leading edge of the dune profile. The shear stress scales as ${\tau}{\sim}{\mu}V_i/{\delta}$,
where ${\mu}$ is the dynamic viscosity of the fluid and ${\delta}$ is a length scale characterizing the inner part of the turbulent boundary layer. In typical
high Reynolds number flow scenarios, the length scale ${\delta}{\sim}0.2{\delta}_{99}$, where ${\delta}_{99}$ is the total turbulent boundary layer
thickness given as ${\delta}_{99}{\sim}{0.37x}{Re_x^{-1/5}}$ \cite{schlichting2016boundary}.
In this work, we study erosion-deposition and particle transport dynamics during the interaction of a turbulent open channel flow with a subaqueous granular structure over a range of Shields number and Froude number as shown in fig. \ref{fig-1}(b).
\section{Materials and Methods}
\label{mat-meth}
The schematic (top view, side view) and the actual experimental setup is shown in fig. \ref{fig-2}(a), \ref{fig-2}(b), and \ref{fig-2}(c) respectively. The experimental setup consists of a closed circuit turbulent open channel flow interacting with a granular subaqueous dune like structure. The open-channel is made out of a horizontal glass-walled flume with dimensions ($1226{\times}137{\times}150$ mm$^3$). The various components are labelled numerically in fig. \ref{fig-2}(a), \ref{fig-2}(b). The flume is fitted with a pipe (1) that delivers water at the inlet section of flume (2). The water from the inlet section flows through a flow conditioner (3) that breaks big eddies into smaller eddies. The flow is then redirected to flow through an inlet gate (4) whose height can be changed to control the flow velocity scale that interacts with the dune. Filtered washed sand particles (sub-angular grains) with size distribution as shown in \ref{fig-2}(d) was used create the dune (5) with an isosceles triangular cross sectional profile with dune width $L{\sim}100mm$, angle of repose of $32^{\circ}{\pm}1^{\circ}$ as initial conditions. The intial heap profile was generated using a custom designed sand dispenser funnel (6) that works underwater. The water height in the flume was controlled by an exit gate (7) and the flowing water out of the flume was redirected using an exit guide duct (8) through a flow strainer (9) to filter particles towards a storage tank (10) with a capacity of $250$ litres. The water from the storage tank is recirculated using a submersible pump (Kirloskar Brothers Limited, KOSN-0520, 0.5 HP, Single Phase, 210 V) (11) through a flow control valve (12) to control the steady state turbulent discharge. The flow rate is measured using a flow meter (Dijiang-OF06ZAT, range: 0-50 LPM, accuracy: 0.5\%) (13) and digital acquisition system using Arduino Uno. The turbulent interaction of the flow with the subaqueous granular structure is characterized using optical diagnostics using white light source (Harison F4A) (14) fitted with a diffuser and a DSLR camera (Nikon, D750) equipped with a lens (Nikon AF-S Nikkor 24-120 mm, 1:4G ED ) providing a spatial resolution of $1280{\times}720$ pixel$^2$ and pixel resolution of $0.1$ mm/pixel at $60$ FPS(frames per second).
All experiments were carried out at atmospheric conditions with a room temperature of $25^{\circ}$C. The flow rate was kept constant at the maximum capacity of the pump at $23.23{\pm}0.08$ LPM (litres per minute) providing an inlet Reynolds number of $Re_i{\sim}2827$.
The flow across the channel
width is mostly uniform apart from very close distance near the
walls (typical of a turbulent flow). The flow at the location where
the flow interacts with the dune is mostly uniform in the absence
of the dune. In the presence of the dune the flow interact with the
dune and changes significantly both upstream and
downstream.
Corresponding to flow rate in the flume, various operating conditions of Shields number (${\theta}$) and Froude number ($Fr$) was generated using a combination of inlet gate and exit gate. The inlet flow velocity $V_i$ was controlled using the inlet gate, whereas the water height ($H_w$) was controlled using the exit gate. For a particular operating condition corresponding to $H_w$, the water height is maintained in hydrostatic condition by closing the exit gate to deploy the particles in a heap.
$370$ gm of sand was deposited using the funnel in a hydrostatic condition to obtain dune height ($H_h$) to water height ($H_w$) ratio of $H_h/H_w{\sim}0.62,{\:}0.69,{\:}0.78$ respectively by changing the exit gate opening. After the particles were deposited in a heap, the funnel was taken out of the flume, the submersible pump was started and the exit gate was opened to maintain a particular $H_w$ for a given operating run. $3-5$ trials were conducted for each operating conditons to assure repeatability and generate the equivalent statistics. All data analysis and visualization were done using ImageJ and in-house python codes. The transient kinematics and shape of subaqueous heap were characterized using an 11-sided polygon vertex method to obtain various shape descriptors like bounding rectangle, and centroid.


\begin{figure}[h]
\centering
\includegraphics[width=6cm,clip]{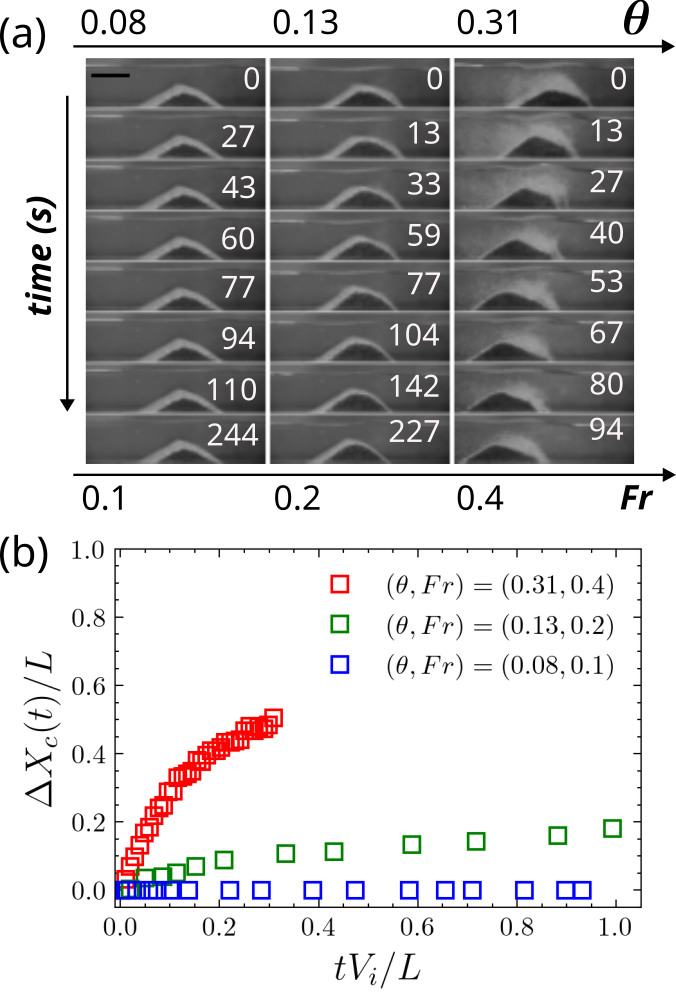}
\caption{(a) Time evolution of dune with Shields and Froude number as a parameter. Note, the mean fluid flow is from right to left. The black scale bar on the left corresponds to 33 mm. (b) The normalized change in dune centroid position as a function of normalized time for various parametric values of ${\theta},{\:}Fr$. }
\label{fig-3}       
\end{figure}

\begin{figure}[h]
\centering
\includegraphics[width=4.5cm,clip]{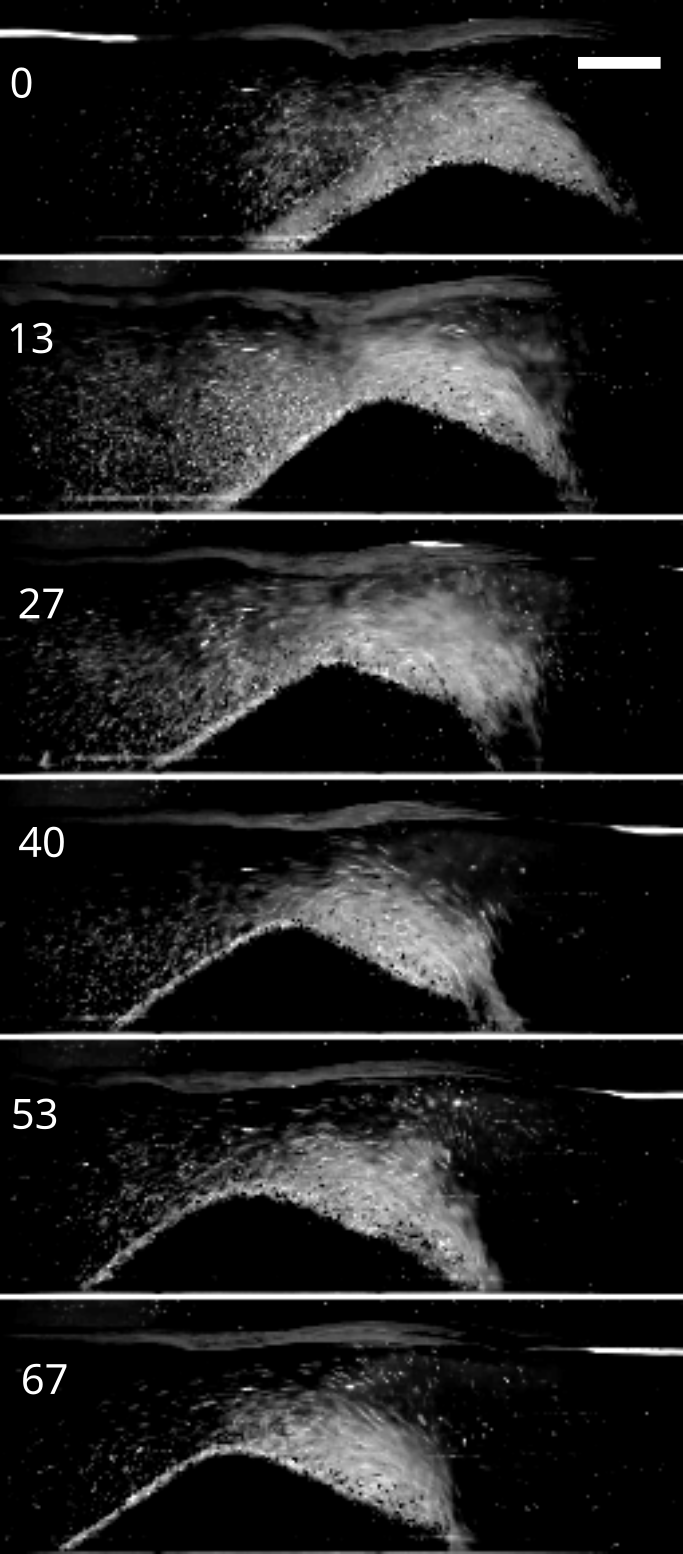}
\caption{Transient erosion and heap particle distribution at $({\theta},Fr)=(0.31,0.4)$. Note, the mean fluid flow is from right to left. The time stamps in white is in seconds. The white scale bar in upper right corresponds to 16 mm.}
\label{fig-4}       
\end{figure}


\section{Results and Discussions}
Fig. \ref{fig-1}(b) shows the experimental regime map in ${\theta}-Fr$ space used in the current experiments. At low values of Shields and Froude number, we observe negligible erosion and particle transport. At intermediate values of Shields and Froude number, erosion starts and the particle transport occurs along the curved surface of the subaqueous structure known as bed load. Bed load charactersitics is associated with particle rolling, sliding and particle trajectories very close to the heap structure. The initiation of particle movement occurs when the the bed shear stress exceeds a critical shear stress (${\tau}_b>{\tau}_{cr}$) value and is characterized by a critical Shields number (${\theta}_{cr}$). From fig. \ref{fig-1}(b) we observe that the scale of
the critical Shields number is ${\theta}_{cr}{\sim}0.1$.
At higher values of Shields and Froude number, the erosion becomes significantly strong and in addition to bed load, suspended load is also observed. Suspended load corresponds to eroded particles getting transported along the mean flow as sediments. However, depending on the particle density ratio and size characterized by particle Stokes number, the particle trajectories in the turbulent flow shows extreme behavior like particle sedimentation, clustering, and tracer-like transport to name a few. The particles gets picked up from the leading edge side and gets deposited downstream. The particle deposition at various locations along the subaqueous structure causes the local slope angle to increase. As the local slope becomes larger than the equilibrium angle of repose ($32^{\circ}\pm1^{\circ}$), particle avalanche starts to occur. The avalanche opposes the fluid flow in the upstream side to move particles downstream on average. The avalanche assists the fluid flow in the downstream direction to transport particles further downstream. The avalanche dynamics in conjuction with the turbulent flow eroding the subaqueous structure results in a motion of the structure along with shape deformation. Fig. \ref{fig-3}(a) shows the transient evolution time series images depicting the motion and subsequent deformation for various values of
Shields and Froude number.
The change in centroid position is normalized with respect to the subaqueous structure length scale $L$, and time $t$ is normalized with respect to kinematic time scale $L/V_i$ (refer to fig. \ref{fig-1}(a)).
Fig. \ref{fig-3}(b) depicts the evolution of normalized x-centroid change in position (${\Delta}X_c(t)/L$) as a function of normalized time ($tV_i/L$).
It can be observed from fig. \ref{fig-3}(a), and fig. \ref{fig-3}(b) that the structure movement and shape deformation is significant at higher values of ${\theta}{\sim}0.31$, and $Fr{\sim}0.4$. The dominant transport mode is both bed and suspended load.
At intermediate values of ${\theta}{\sim}0.13$, and $Fr{\sim}0.2$, the structure moves at very slow speed as a rigid body and structure shape remains almost invariant in time.
The dominant transport mode is bed-load.
At lower values of ${\theta}{\sim}0.08$, and $Fr{\sim}0.1$ we observe no erosion, structure movement and deformation. Fig. \ref{fig-4} shows the particle distribution in the flow for high values of ${\theta}{\sim}0.31$, and $Fr{\sim}0.4$. The transient shape evolution and the centroid motion towards the left can also be observed clearly in fig. \ref{fig-4} (Note, the mean fluid flow is from right to left). We can also observe from fig. \ref{fig-4}, the presence of vortical structure at the upstream side of the subaqueous structure. The vortex causes the upstream side of the dune to form an avalanche face due to particle deposition and shear stress induced by a counter-clockwise vortex.

\section{Conclusion}
In conclusion, we study the erosion-deposition, centroid kinematics and shape deformation of a subaqueous sand dune-like structure for various parametric values of Shields and Froude number using high-fidelity optical diagnostics. We show that, at low values of Shields and Froude number there is no erosion, structure motion and shape deformation. At intermediate values of Shields and Froude number, we observe slow erosion, particle transport in the form of bed-load and very slow motion of the structure centroid with negligible shape deformation. At high values of of Shields and Froude number, erosion is significant with the dominant transport being a combination of bed-load and suspended load. The centroid of the subaqueous structure moves significantly and the corresponding shape also evolves as a result of a turbulent flow particle interactions at various spatio-temporal scales. We also unearthed the presence of a vortex at the upstream side of the dune that assists in the formation of avalanche faces.
\bibliography{ref}
\end{document}